\documentclass[useAMS,usenatbib]{mn2e}
\usepackage{amsmath}
\usepackage{amssymb}
\usepackage{bm}
\usepackage{dcolumn}
\usepackage{graphicx}
\usepackage{color}
\usepackage{ulem}

\newcommand       \apj          {ApJ}
\newcommand       \apjl         {ApJL}
\newcommand       \aap          {A\&A}
\newcommand       \nat          {Nature}
\newcommand       \mnras        {MNRAS}

\newcommand       \prd      {Phys.~Rev.~D.~}
\newcommand       \araa      {ARA\&A}

\newcommand      \physrep {Phys.~Rep.}

\makeatletter
\def\tbcaption{\def\@captype{table}\caption}
\def\figcaption{\def\@captype{figure}\caption}
\makeatother

\title[Radioactive decay products in NS merger ejecta]
  {Radioactive decay products in neutron star merger
  ejecta: heating efficiency and $\gamma$-ray emission}
\author[K.~Hotokezaka et al.]
  {K. Hotokezaka\thanks{E-mail: kenta.hotokezaka@mail.huji.ac.il}$^1$,
  S. Wanajo$^{2, 3}$, 
M. Tanaka$^4$, 
A. Bamba$^5$, 
 Y. Terada$^6$, 
and  T. Piran$^1$
\\
  $^1$Racah Institute of Physics,~The Hebrew~University of Jerusalem,~Jerusalem~91904,~Israel.\\
  $^2$Department of Engineering and Applied Sciences, Sophia University, Chiyoda-ku, Tokyo 102-8554, Japan.\\
  $^3$iTHES Research Group, RIKEN, Wako, Saitama 351-0198, Japan.\\
  $^4$National Astronomical Observatory of Japan, Mitaka, Tokyo 181-8588, Japan.\\
  $^5$Department of Physics and Mathematics, Aoyama Gakuin University 5-10-1 Fuchinobe Chuo-ku, Sagamihara, Kanagawa 252-5258, Japan.\\
  $^6$Graduate School of Science and Engineering, Saitama University, 255 Shimo-Okubo, Sakawa, Saitama, Saitama 338-8570, Japan.\\
  }

\date{17 Nov, 2015}

\pagerange{\pageref{firstpage}--\pageref{lastpage}} \pubyear{2015}

\begin{document}

\label{firstpage}

\maketitle

\begin{abstract}
The radioactive decay of the freshly synthesized $r$-process nuclei ejected in compact binary mergers power 
optical/infrared macronovae (kilonovae) that follow these events.  The light curves depend critically 
on the energy partition among the different products of the radioactive decay and this plays an important 
role in estimates of the amount of ejected $r$-process elements from a given observed signal. 
We study the energy partition and $\gamma$-ray emission of the radioactive decay.
We show that $20$--$50\%$ of the total radioactive energy
 is released in $\gamma$-rays on
timescales from hours to a month. The number of emitted $\gamma$-rays
per unit energy interval has roughly a flat spectrum between a
few dozen keV  and $1$~MeV so that most of this energy
is carried by   $\sim 1$~MeV $\gamma$-rays. However at the peak  of macronova emission the optical depth of 
the $\gamma$-rays is $\sim 0.02$ and  most of the $\gamma$-rays escape. 
The loss of these $\gamma$-rays reduces the heat deposition into the ejecta and hence reduces the 
expected macronova signals if those are lanthanides dominated. This implies that the 
ejected mass is larger by a factor of $2$~--~$3$ than what was previously estimated.
Spontaneous fission heats up the ejecta and
the heating rate can increase if a sufficient amount of transuranic nuclei are synthesized. 
Direct measurements of these escaping $\gamma$-rays may provide the ultimate proof for the 
macronova mechanisms and an identification of the $r$-process nucleosynthesis sites. 
However, the chances to detect these signals are slim with current X-ray and $\gamma$-ray missions.
New detectors, more sensitive by at least a factor of ten, are needed for a realistic detection rate.
\end{abstract}


\maketitle

\section{Introduction}
\label{sec:Introduction} The origin of a half of the elements heavier than iron such as
gold and uranium, which are made by the $r$(apid)-process, is one
of the current nucleosynthesis long-standing
mysteries~\citep{cowan1991PhR,wanajo2006NuPhA,qian2007PhR,arnould2007PhR}.
Compact binary mergers (neutron star--neutron star or neutron
star--black hole binaries) that  eject highly neutron rich material in which
$r$-process nucleosynthesis would naturally take place have been suggested as promising production sites~\citep{lattimer1976ApJ,symbalisty1982ApL,eichler1989Nature,freiburghaus1999ApJ}.

While the overall amount of heavy r-process material in the Galaxy is consistent with 
expectations of mass ejection in mergers and with their expected 
rates~(see, e.g., \citealt{kalogera2004ApJ,kim2015MNRAS,wanderman2015MNRAS} 
for the merger rate estimates), \cite{argast2004A&A} pointed out a
difficulty for neutron star mergers to reproduce the $r$-element enrichment
for halo stars with very low metallicities due to the delay time of the merger events.
This difficulty can be resolved  by taking into account 
the turbulent mixing of material in the galaxy~\citep{piran2014}, the  assembly of
sub-halos during the formation of the Galaxy~\citep{vandevoort2015MNRAS,ishimaru2015ApJ,shen2015ApJ,hirai2015} 
or if the delay time of mergers in the early Universe is shorter than what current models suggest.
The large scatter in the abundances of $r$-process elements of metal poor stars
can be naturally explained by the rarity of $r$-processing events such as 
neutron star mergers~(e.g., \citealt{tsujimoto2014A&A,wehmeyer2015MNRAS}). 
In addition, \cite{hotokezaka2015} have shown that the rarity of the merger events is 
broadly consistent with the $^{244}$Pu abundances of the early solar
system material~\citep{turner2007E&PSL} and the present-day deep-sea archives~\citep{Wallner2015NatCo}.

The merger origin scenario of  $r$-process elements
has been attracting even more attention since the discovery of a possible
macronova, also called as a kilonova, associated with the short gamma-ray 
burst~(GRB)~130603B~\citep{tanvir2013Nature,berger2013ApJ}. 
More recently, it
has been shown that there is an infrared excess in the
afterglow data of GRB~060614~\citep{yang2015NatCo} as well, 
suggesting another macronova candidate.  A maronova is a radioactively
powered transient  as an electromagnetic
signal of $r$-process elements in merger ejecta
\citep{li1998ApJ,metzger2010MNRAS,roberts2011ApJ,
korobkin2012MNRAS,barnes2013ApJ,
tanaka2013ApJ,grossman2014MNRAS,kasen2015MNRAS}. While both candidates
are based on single data points, the observed data are
largely consistent with theoretical expectations of macronovae and, if
correct, it is a strong evidence that  compact binary
mergers eject significant amounts of $r$-process
material.  The association of these macronovae with short GRBs provide the first direct evidence, supporting 
the  circumstantial evidence~\citep{nakar2007,berger2014}, that 
compact binary mergers are the progenitors of short gamma-ray bursts~\citep{eichler1989Nature}.

The minimal required mass to explain the brightness
of the macronova candidate associated with GRB~130603B is estimated as
$0.02~M_{\odot}$~\citep{hotokezaka2013ApJL,piran2014}. This is
consistent with recent hydrodynamical simulations
of compact binary mergers
\citep{hotokezaka2013PRDa,bauswein2013ApJa,rosswog2013RSPTA,
foucart2013PRD,fernandez2013MNRAS,
metzger2014MNRAS,perego2014MNRAS,just2015MNRAS,
kyutoku2015,sekiguchi2015PRD,kawaguchi2015PRD,kiuchi2015PRD,east2015}. The minimal required mass
for the macronova candidate associated with GRB~060614 is even larger of order $0.1 M_\odot$
\citep{yang2015NatCo} and 
this can possibly be explained only in  a neutron star-black hole 
merger~\citep{foucart2013PRD,just2015MNRAS,kyutoku2015,kiuchi2015PRD}.

Estimates of the mass of ejected $r$-process elements depend not only on
the overall energy generation rate of the radioactive decay but also on
the efficiency that the energy of emitted decay products is deposited in the ejecta.
While the overall energy generation rate has been extensively studied~\citep{metzger2010MNRAS,roberts2011ApJ, korobkin2012MNRAS,
wanajo2014ApJ,lippuner2015}, the heating efficiency has not been well studied.
In the literature, it is often assumed that  $30$--$50\%$ of the generated energy is deposited in the ejecta.
In order to evaluate the efficiency, one needs to identify the
energies released in different
decay products: electrons, neutrinos, $\gamma$-rays,
fission fragments, and $\alpha$-particles.  Here we determine the energy
partition  among the different types of products and estimate the net energy deposition rate  in the ejecta.
We pay particular attention to the escape of $\gamma$-rays from the ejecta at the
stage when it is still optically thick to IR/optical/UV radiation. 

In addition, it is worthwhile   to study the spectrum and flux of the
$\gamma$-rays escaping from the macronova. 
 The detection
of $\gamma$-ray lines of specific heavy nuclei can 
 provide a conclusive evidence for 
 $r$-processing during a merger event.  While the detection of
$\gamma$-ray lines of radioactive decay from astrophysical sources
is quite challenging, they
have been detected from nearby supernovae: the $\gamma$-ray
lines of $^{56}$Co from the type II SN~1987A
\citep{matz1988Natur,teegarden1989Natur} and 
those of $^{56}$Ni and $^{56}$Co from the type Ia
SN~2014J~\citep{diehl2014Sci,churazov2014Natur,terada2015ApJ}. 
Due to the rarity of macronova this detection is even more challenging.
Still it is interesting to  discuss the
expected observational features of $\gamma$-ray lines of
$r$-process nuclides from neutron star mergers and their
detectability with  current and future X-ray and
$\gamma$-ray facilities.

\section{Model set up}
The abundance pattern of synthesized nuclei in 
merger ejecta depends on the fluid element's initial state, expansion
velocity, and neutrino irradiation from the central remnant object 
\citep{freiburghaus1999ApJ,metzger2010MNRAS,roberts2011ApJ,korobkin2012MNRAS,wanajo2014ApJ,goriely2015MNRAS}.
Although the resulting patterns obtained from
such merger models are not necessarily
similar to the solar $r$-process abundance pattern,
it is known that the abundances of $r$-process elements in
$r$-process-enhanced stars in the Galactic halo closely follow
the solar $r$-process one, in
particular for $Z > 50$ with almost perfect agreement (see, e.g.,
\citealt{sneden2008ARA,barbuy2011A&A,roederer2014ApJ,siqueira2014A&A}).  This
fact suggests that  a single phenomena 
reproduces the solar-like $r$-process abundance patterns.

We assume \citep[see][]{tanaka2014ApJ}, therefore, that the nuclear abundance distribution ($Y_A$, in
the range $90 \leq A \leq 238$ for the fiducial case, see below) that
matches the solar pattern of stable and long-lived ($^{232}$Th and
$^{235, 238}$U) $r$-process nuclei \citep{cowan1999ApJ}. 
Nuclear abundances ($Y_A = \sum_Z Y_{Z, A})$ are thus
time-independent while isotopic abundances ($Y_{Z, A}$) are
time-dependent. The initial $Z$ for each $Y_{Z, A}$ is set at the neutron
separation energy of about 2~MeV (roughly at the $r$-process freezeout)
in the very neutron-rich side of the chart of nuclides. The time
evolution of all isotopes is then calculated by a reaction network
code described in \citet{wanajo2014ApJ}. All the reaction channels
except for $\beta$-decays (that do not change $Y_A$) are switched
off~(for $A<206$, see below). Uncertainties in theoretical estimates of $\beta$-decay lifetimes are
irrelevant here because most of the isotopes decay back to the vicinity
of $\beta$-stability after several seconds, where experimental
half-lives and $Q_\beta$-values are available. Then we determine the
energy partition into each type of decay product and the resulting
$\gamma$-ray spectra. Here we use the
nuclear data base Evaluated Nuclear Data File ENDF/B-VII.1
library~\footnote{https://www-nds.iaea.org/public/download-endf/ENDF-B-VII.1/decay/}
for electrons, neutrinos, and $\gamma$-rays.

\begin{figure*}
\includegraphics[width=80mm]{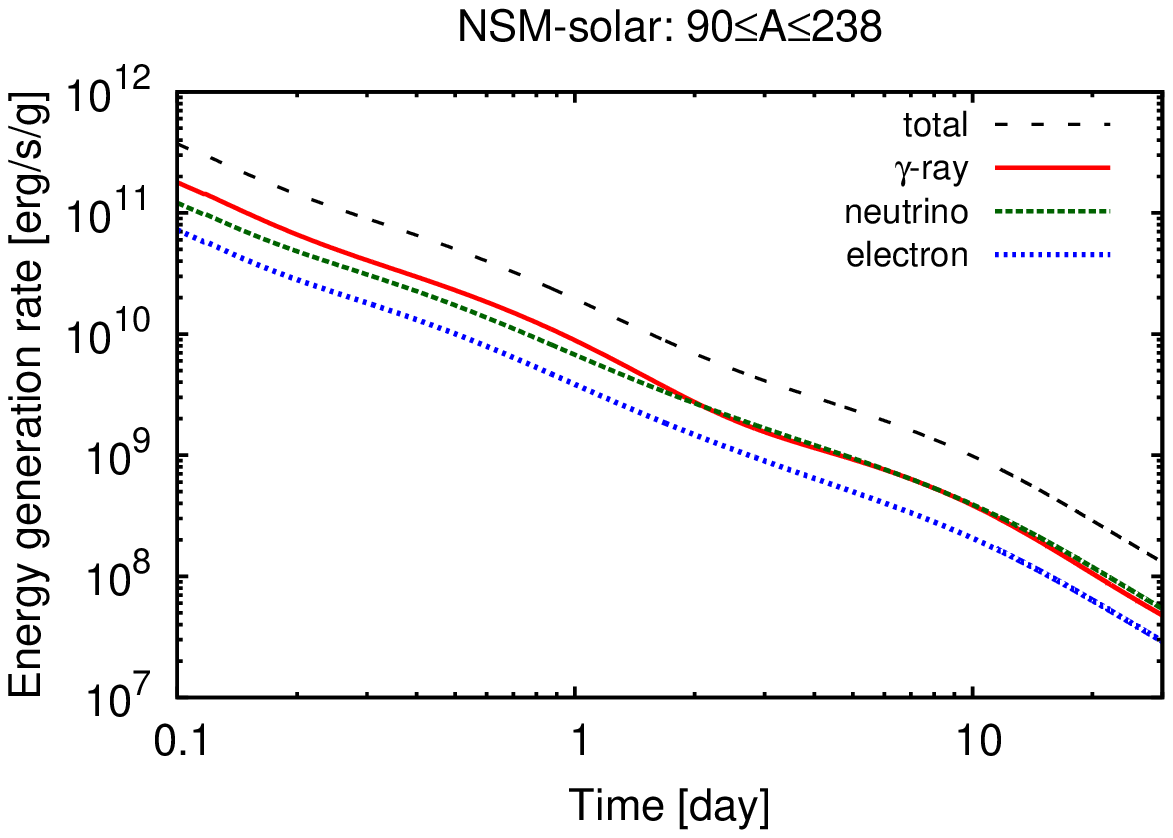}
\includegraphics[width=80mm]{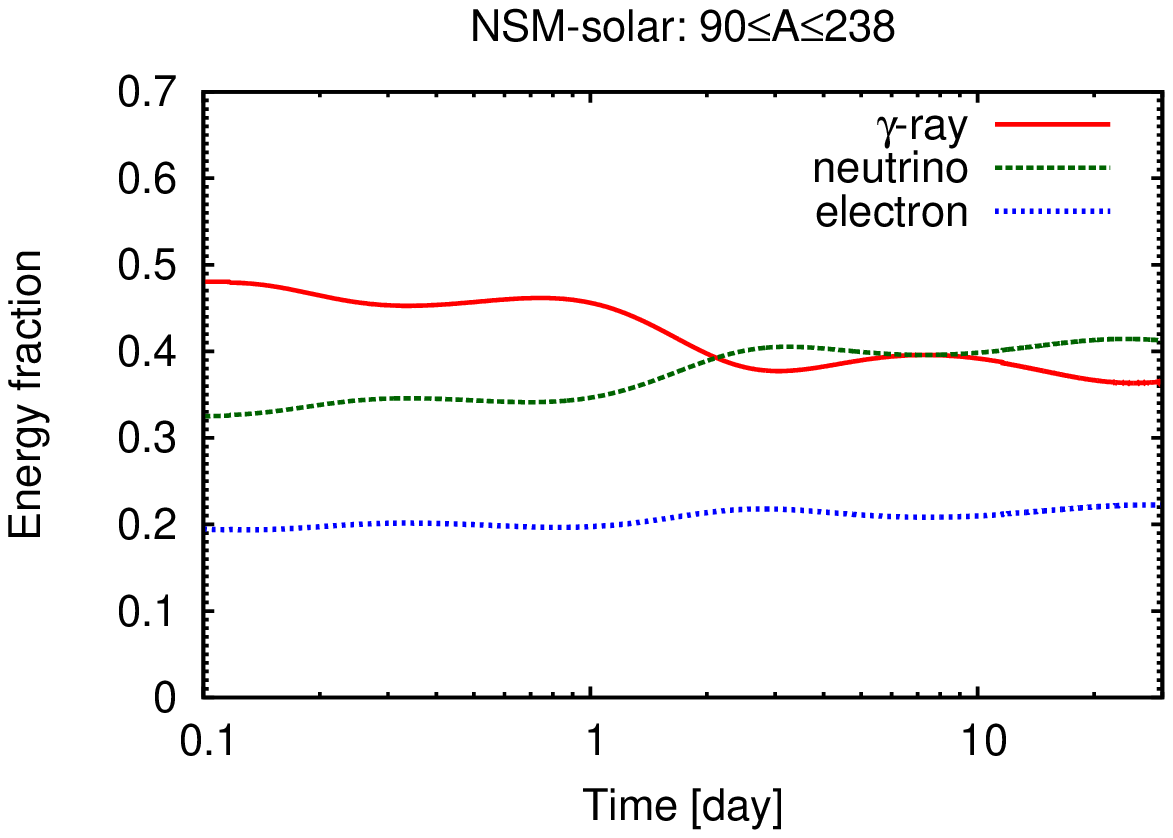}\\
\includegraphics[width=80mm]{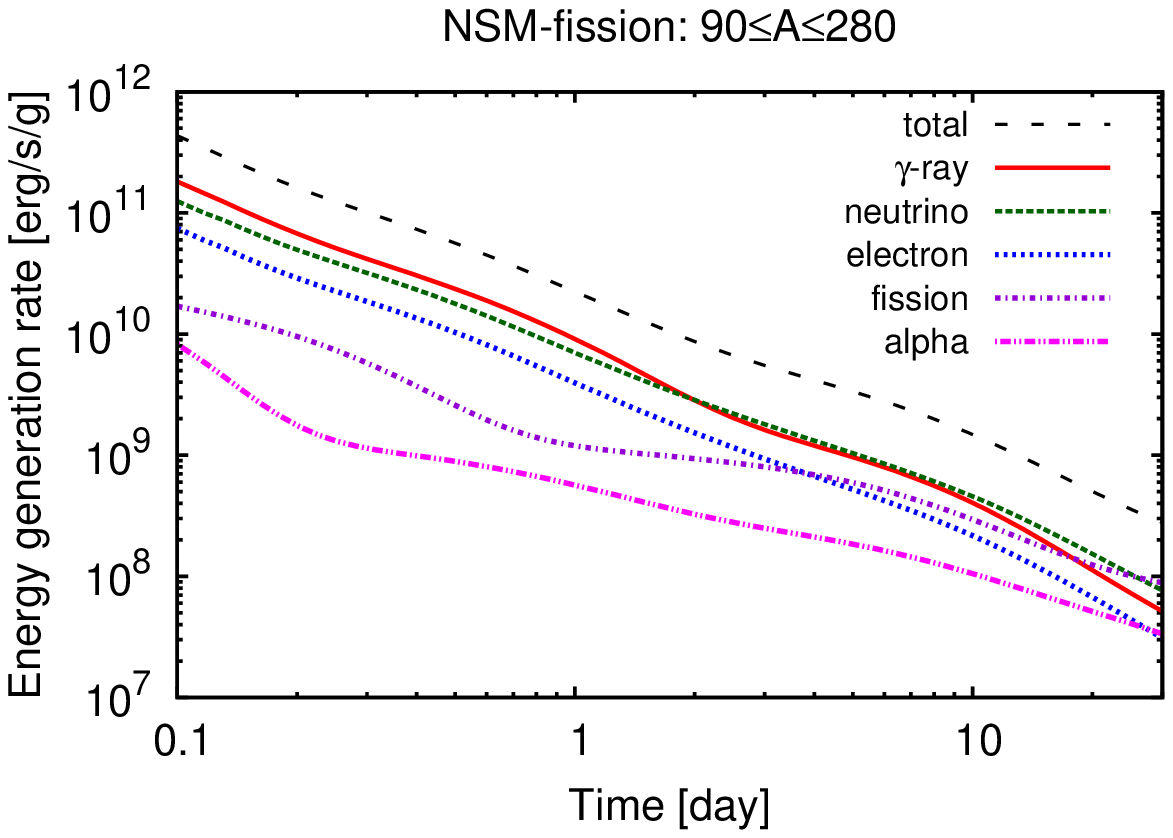}
\includegraphics[width=80mm]{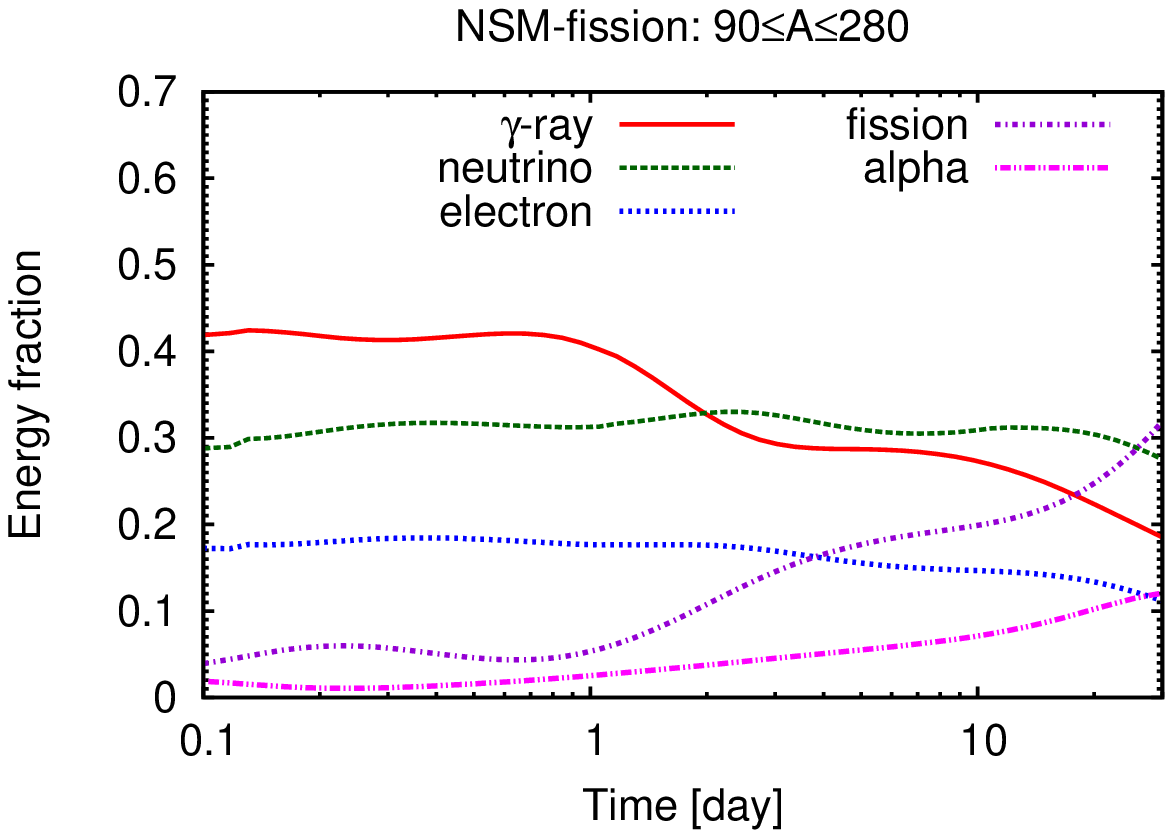}\\
\includegraphics[width=80mm]{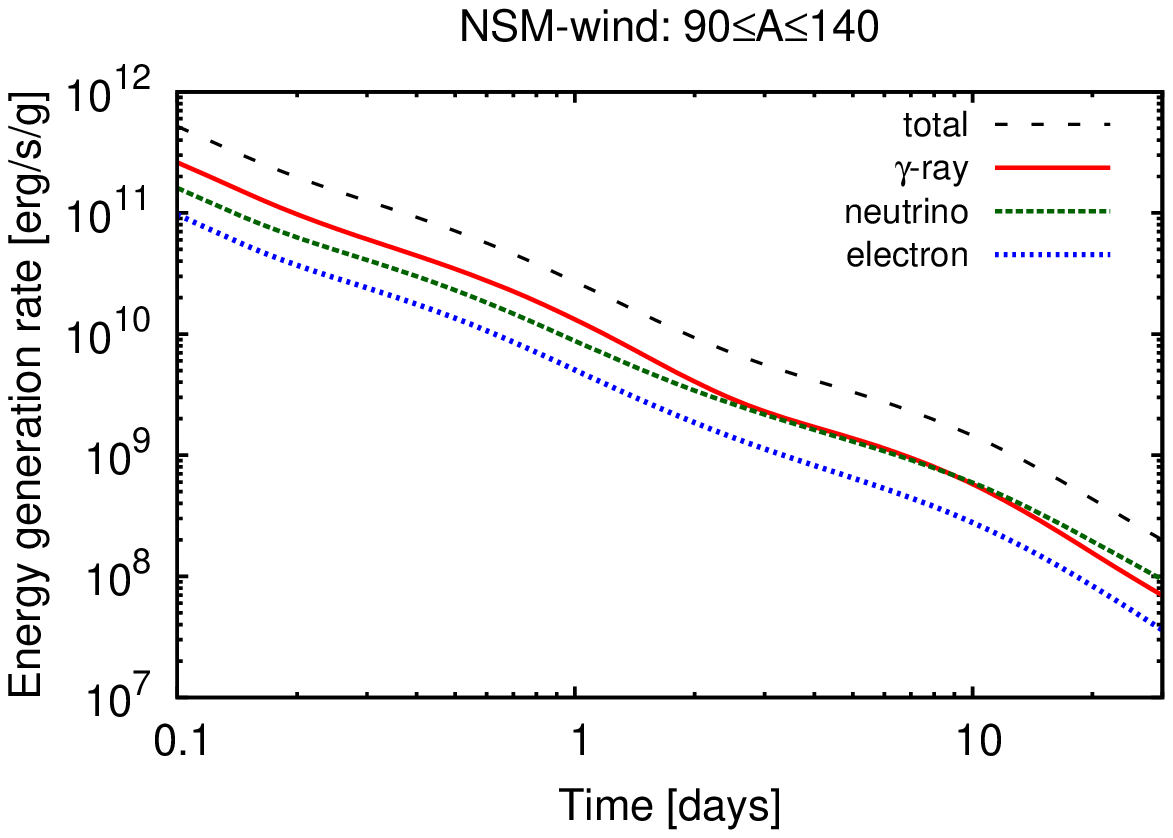}
\includegraphics[width=80mm]{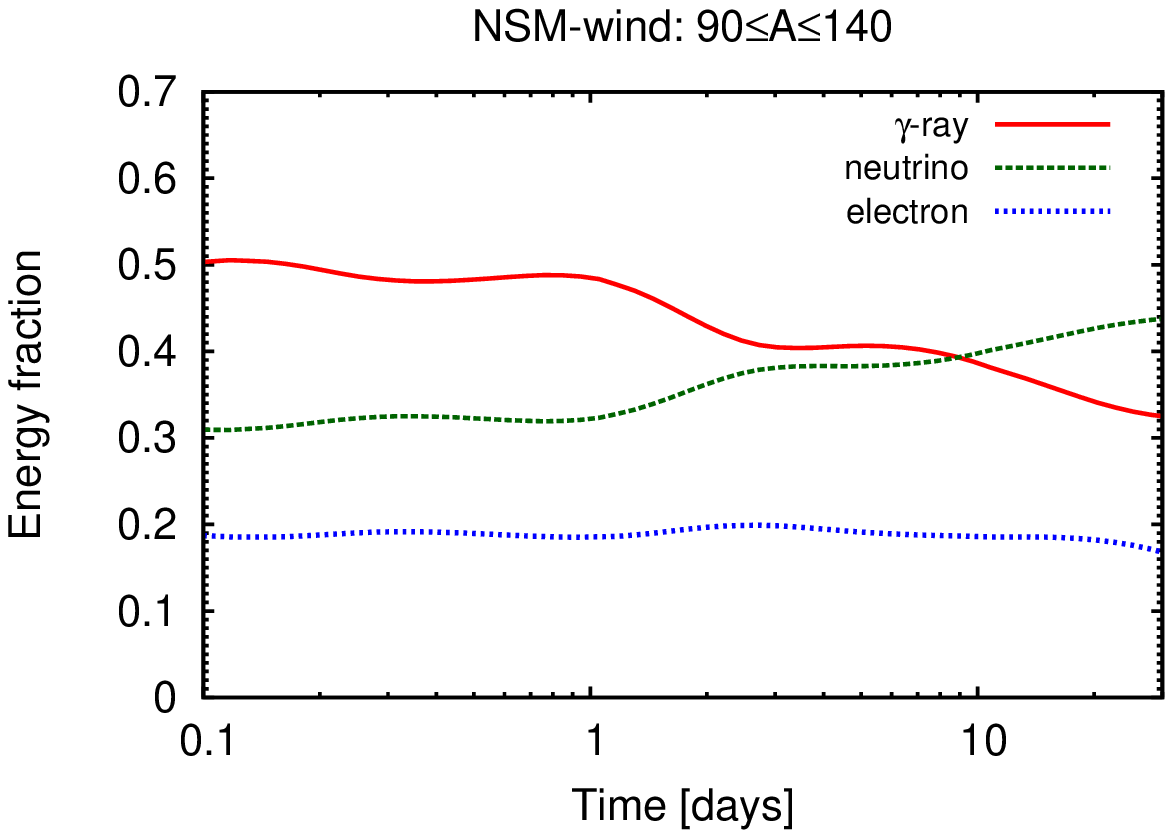}\\
\caption{Energy generation rate in each type of particles~(left) and
 its fraction to the total 
one~(right) for NSM-solar~($90\leq A \leq 238$),
NSM-fission~($90\leq A \leq 280$), 
and NSM-wind~($90\leq A \leq 140$) 
from the top to the bottom.
Each curve shows the total rate~(black long-dashed), those in the
forms of $\gamma$-rays~(red solid), neutrino~(green dashed), electrons~(blue dotted), 
fission fragments~(violet dash-dotted), and $\alpha$ particles~(magenta dash-two dotted).
}
\label{fig:fraction}
\end{figure*}

Although there are no stable~(long-lived) nuclides with mass numbers above $A=238$ on
Earth, some short-lived transuranic
nuclei with $A>238$ are likely produced by the $r$-process. For example, 
there is clear evidence that $^{244}$Pu
and $^{247}$Cm (with the
half-lives of 81~Myr and 16~Myr, respectively) existed in the early
solar system~\citep{turner2004Sci,brennecka2010Sci} and $^{244}$Pu is found in
the Earth's material at present~\citep{Wallner2015NatCo}.  Furthermore,
nucleosynthesis studies of merger ejecta show that very heavy
nuclei up to mass numbers of $\sim 280$ 
exist at the $r$-process freezeout~\citep[see,
e.g.,][]{goriely2013PRL,eichler2015ApJ}.  The spontaneous fission of such very
heavy nuclei is also suggested to affect the
heating rate~\citep{metzger2010MNRAS,wanajo2014ApJ}. In this work,
we study  three cases: $r$-process nuclear
distributions of
(i) NSM~(Neutron Star Merger)-solar: $90\leq A \leq 238$~(fiducial), 
(ii) NSM-fission: $90\leq A\leq 280$, and
(iii) NSM-wind: $90\leq A \leq 140$.
The last case, NSM-wind corresponds to the conditions within 
 a possible lanthanide-free composition
(from the wind, see below).  For NSM-fission, 
we add the transuranic nuclei by assuming a constant
$Y_A$'s of $3.6\cdot 10^{-4}$ for
$206\leq A\leq 280$. These values are taken so
that the solar abundance of $^{209}$Bi is reproduced after nuclear
decay. Note that the bulk of $^{206, 207,208}$Pb, $^{209}$Bi, $^{232}$Th, 
and $^{235, 238}$U are the ($\alpha$
and $\beta$) decayed products of actinides with $209 < A < 254$.
The reaction network includes the channels for
($\beta$-delayed and spontaneous) fission and $\alpha$-decay in addition
to $\beta$-decay for this mass region.

To study the heating efficiencies and resulting $\gamma$-ray line fluxes, 
one needs to specify the ejecta properties, e.g., the mass $M_{\rm ej}$ and expansion
velocity $v$. In this work, we consider two types of merger
ejecta: dynamical ejecta induced by tidal torque and shock 
heating during the merger~\citep{hotokezaka2013PRDa,bauswein2013ApJa,rosswog2013RSPTA,
foucart2013PRD,kyutoku2015,sekiguchi2015PRD,kawaguchi2015PRD,east2015} and
a wind ejected by neutrino heating,
viscosity, magnetic field, and nuclear recombination from the remnant central object
and its accretion disk~\citep{shibata2011ApJ,wanajo2012ApJ,fernandez2013MNRAS,
metzger2014MNRAS,perego2014MNRAS,siegel2014ApJ,just2015MNRAS,kiuchi2015PRD}.  
We set  as canonical parameters
$(M_{\rm ej},~v)=(0.01M_{\odot},~0.3c)$ for the dynamical ejecta and
$(0.01M_{\odot},~0.05c)$ for the wind.  We also consider an
ejecta mass of $0.1M_{\odot}$ as an extreme case.  Such a large mass
ejection may take place at a black-hole neutron star
merger~\citep{foucart2013PRD,just2015MNRAS,kyutoku2015,kiuchi2015PRD}.

\section{Energy partition of radioactive decay into different types of
 products}

Decay channels of freshly synthesized $r$-process nuclei
are divided into the
three types: $\beta$-decay, 
$\alpha$-decay, and spontaneous fission\footnote{Other
channels such as neutron-induced fission, $\beta$-delayed fission, and
$\beta$-delayed neutron emission play roles during or immediately after
the $r$-process, which are irrelevant to macronova emission (some days
after the merging).}. Different types of
decay produce  different types of products with 
different amounts of energy.  A single
$\beta$-decay emits one electron and one neutrino
at energies of $\sim 0.1$--$10$~MeV.  In addition to
$\beta$-decay, heavy
nuclei with $A \ge 210$ are unstable to $\alpha$-decay
or to spontaneous fission since the binding energy of such a 
nucleus appreciably decreases with the proton number due to the
Coulomb repulsion. For both types of decay, the binding energy
difference between the initial and final state of an
emitted particle is about a few MeV per nucleon.  Therefore 
a single decay releases about $5$~MeV in kinetic energy for
$\alpha$-decay and $200$~MeV for spontaneous fission,
respectively.  Furthermore, all  types of decay often
leave excited nuclei that emit $\gamma$-rays as they transit to
the ground states.  The timescales of the transitions are much shorter than those in which  we
 are interested in here. The energies of emitted
$\gamma$-rays, which correspond to the differences between
the energy levels of nuclides, range from a few dozen keV
to a few MeV.

It is important to note that a  single spontaneous fission process releases a huge amount of
energy that is comparable  to the energy released by 
$\beta$-decay of the sum of $10$--$100$ nuclei.  It is also worthy
to note the time evolution of energy generation rate 
for each type of decay. For $\alpha$-decay and
spontaneous fission, the energy generation rate evolves as $\propto
t^{-1}$ since each unstable nucleus releases roughly
a fixed amount of energy on the timescale of its
lifetime. On the contrary, $\beta$-decay generates
energy following roughly $\propto t^{-1.3}$.  This is
because for $\beta$-decay the
nucleus with a longer lifetime releases a smaller amount of
energy. Thus, depending on the abundance of transuranic elements,
spontaneous fission potentially plays an important role in the macronova
heating process at late times.

Figure~\ref{fig:fraction} depicts the energy generation rates in
different types of products resulting from $r$-process
material~(left).
Roughly speaking, the energy
generation rates in the forms of $\gamma$-rays~(red solid
curve), neutrinos~(green dashed curve), and electrons~(blue dotted
curve) follow a power law of $t^{-1.3}$. For NSM-fission, the energy
generation rates in spontaneous fission~(violet dash-dotted curve)
and $\alpha$-decay (magenta dash-two dotted curve)
follows roughly $\propto t^{-1}$ as anticipated.
After around $5$ days, the energy released in
the form of fission fragments exceed
the energy released in electrons.  The right panels of
Fig.~\ref{fig:fraction} show the energy fractions released in
all types of decay products.  The energy
fractions of neutrinos and $\gamma$-rays are $0.25$--$0.4$ and
$0.2$--$0.5$, respectively.  
For NSM-solar and NSM-wind, the energy fraction of electrons is $\sim 0.2$ and  
almost constant with time.
For NSM-fission, the energy fraction of electrons 
slowly decreases with time from $0.2$ to $0.1$.  After a few days,
spontaneous fission releases $10$--$35\%$ of the total radioactive
energy.  
Note that the energy fraction of electrons is rather small
compared to the typical value of $\beta$-decay. 
This  small fraction  is resulted from
properties of nuclides around the second $r$-process peak $A \sim
130$. Some of them have larger 
$Q_\beta$-values and emit $\gamma$-rays  at higher
energies.

\begin{figure*}
\includegraphics[width=80mm]{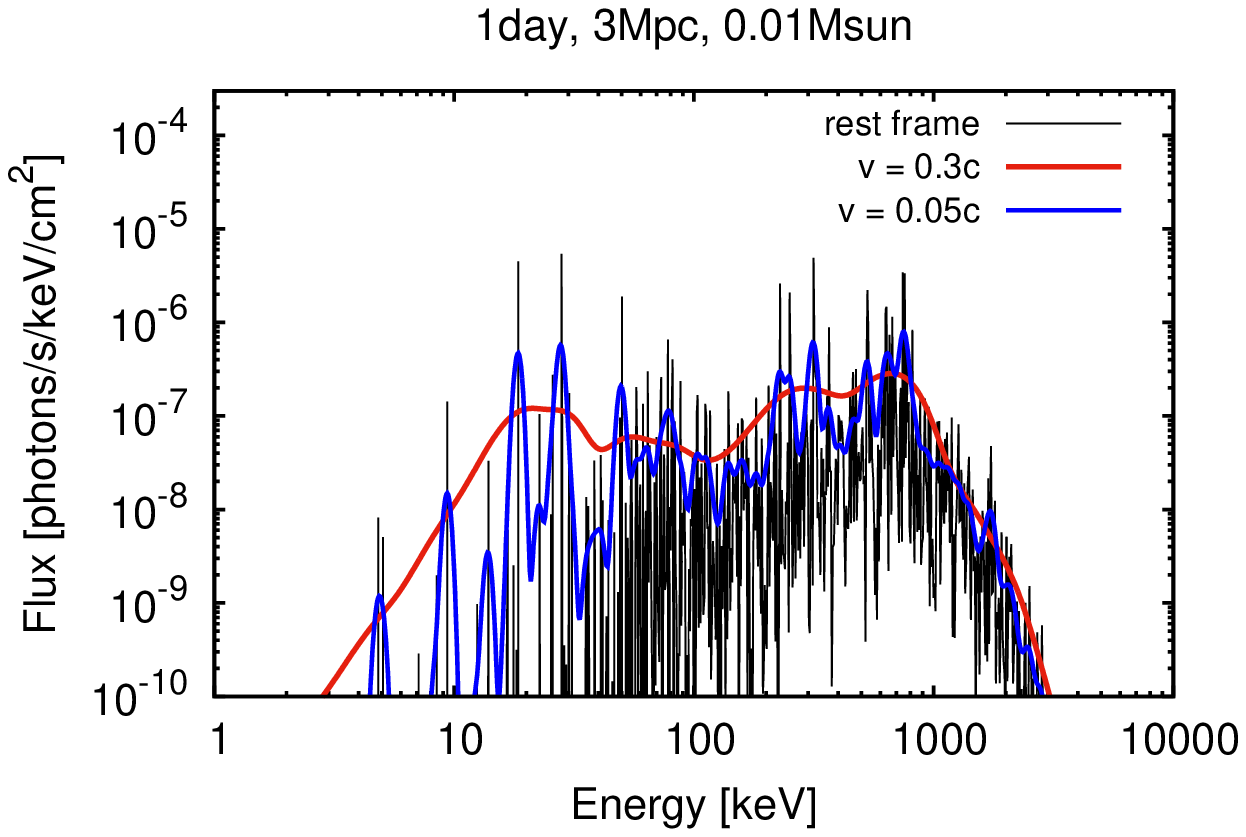}
\includegraphics[width=80mm]{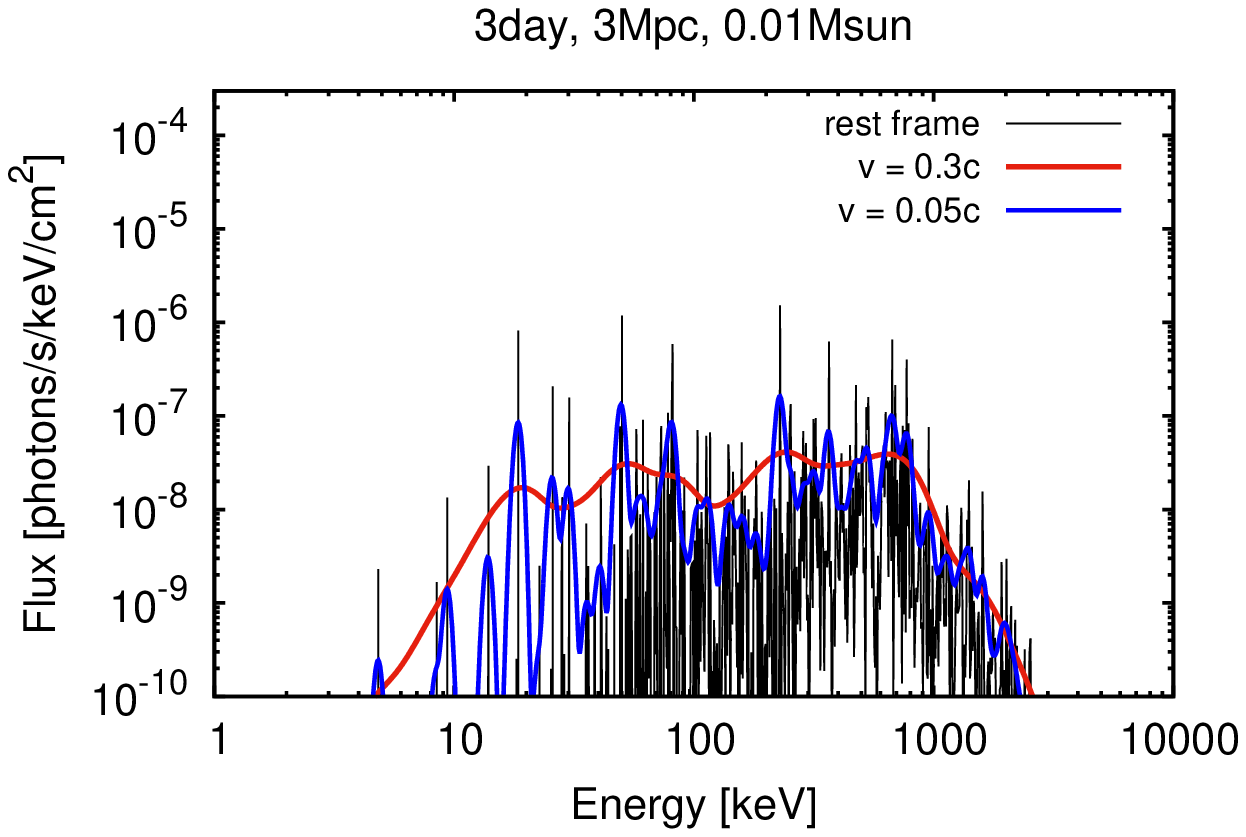}\\
\includegraphics[width=80mm]{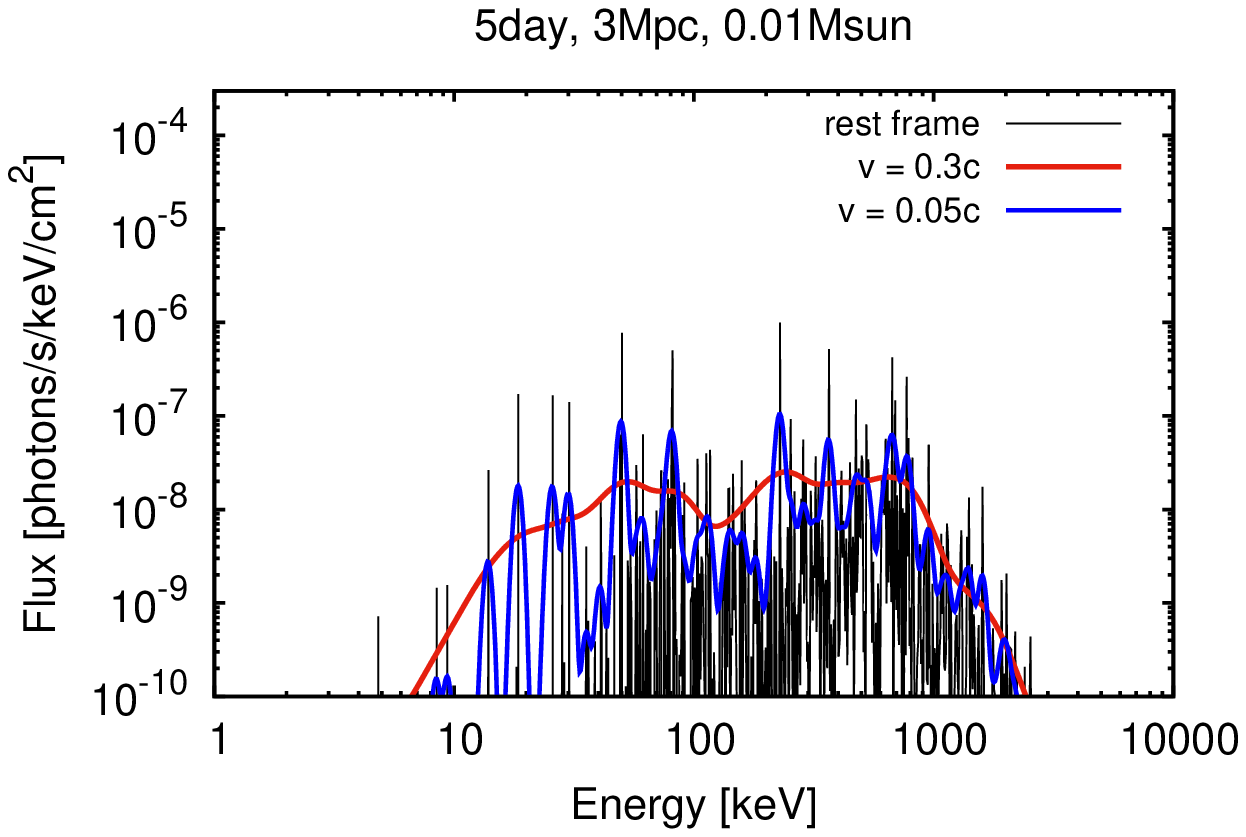}
\includegraphics[width=80mm]{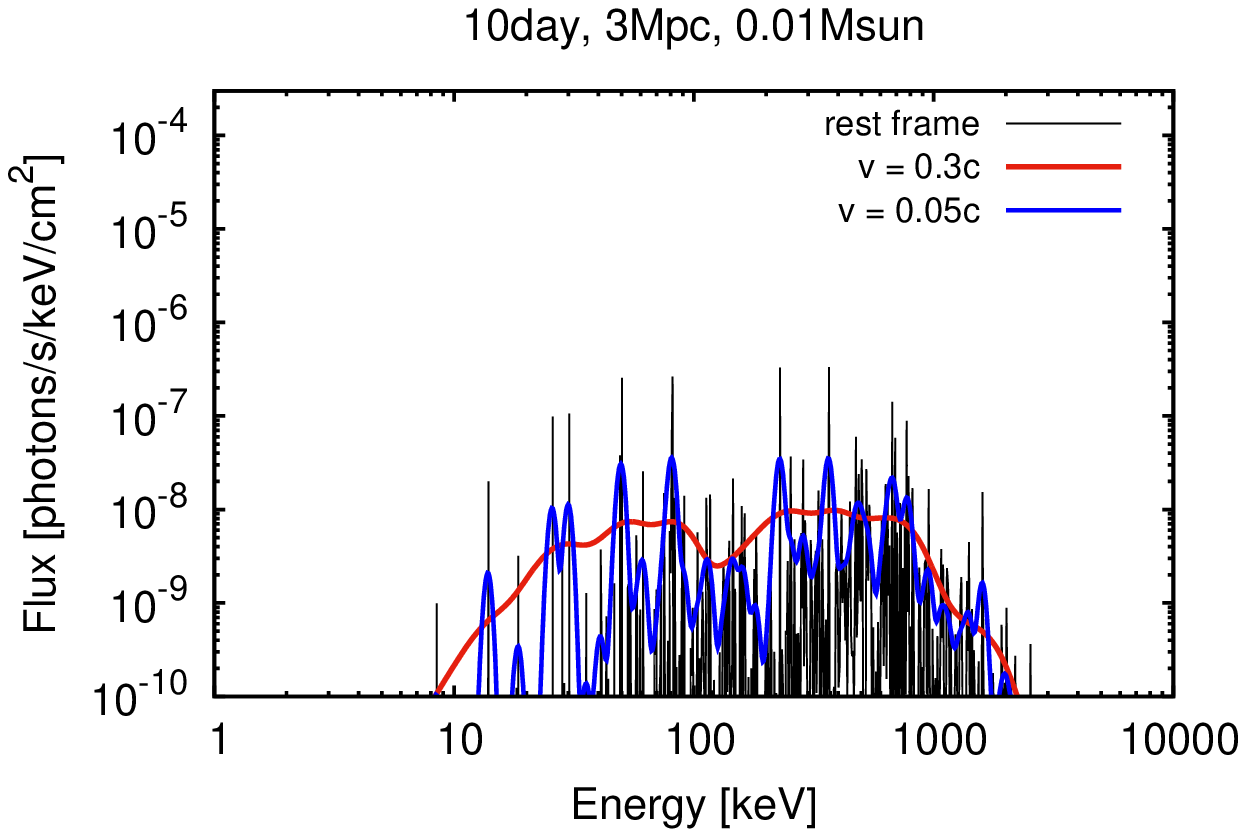}
\caption{Spectrum of $\gamma$-rays at $1$, $3$, $5$
and $10$~days after merger for NSM-solar. Black lines depict
the $\gamma$-ray spectrum produced by nuclei at rest.
The red~(blue) curve shows the spectrum with the Doppler broadening
with an expansion velocity of $0.3c$~($0.05c$). The normalization is determined
with the mass of ejected $r$-process elements of $0.01M_{\odot}$
and the observed distance of $3$~Mpc. Here we do not take
any absorption and scattering processes into account.}
\label{fig:spectrum}
\end{figure*}

In Fig.~\ref{fig:spectrum}, we show the $\gamma$-ray spectra for NSM-solar
at $1$~day, $3$~days,  $5$~days, and $10$~days after the merger. The black lines
in each panel corresponds to the $\gamma$-ray spectrum at
the fluid rest frame. The red~(blue) curve is a spectrum taking into account
 the Doppler effect with an expansion velocity of $v=0.3c$~($0.05c$).
Here each line is convolved with a Gaussian  with a full-width-half-maximum of $2\sqrt{\ln 2}\,v/c$.
The spectra are normalized with an ejecta
mass of $r$-process elements of $0.01M_{\odot}$ for an observer at
a distance of $3$~Mpc. The spectral shape is
approximately flat from a few dozen keV to a few MeV and most of
the $\gamma$-rays' energy  is carried by photons at energies of
$\sim1$~MeV.   The spectrum is not significantly different among the different
models as long as the second $r$-process peak~($A\sim 130$) is contained as in our models.

On the timescales of interest for us, neutrinos completely escape from the
ejecta without any interaction.  Some of the $\gamma$-rays escape as well.  
Electrons, $\alpha$-particles, and fission fragments interact with 
charged particles via Coulomb collision and their
energies may be deposited efficiently in the ejecta.  The
radioactive heating rate of the ejecta $\dot{Q}$ can be written as
\begin{eqnarray}
\dot{Q}(t)  & = &\dot{E}_{e}(t) + \epsilon_{\gamma}(t)\dot{E}_{\gamma}(t) + \dot{E}_{\alpha}(t) + \dot{E}_{f}(t),
\label{Eq.array}
\end{eqnarray}
where
$\dot{E}_{e}$, $\dot{E}_{\gamma}$, $\dot{E}_{\alpha}$, and
$\dot{E}_{f}$ are the energy generation rates in the forms
of electrons, $\gamma$-rays, $\alpha$-particles, and fission
fragments, respectively. We assume all the energy of charged 
decay products are deposited in the ejecta.
For NSM-fission, we find
\begin{eqnarray}
\dot{E}_{e}(t) & \approx & 4\cdot 10^{9}~{\rm erg/s/g}~\left(\frac{t}{1~{\rm day}}\right)^{-1.3},\\
\dot{E}_{\gamma}(t) & \approx & 8\cdot 10^{9}~{\rm erg/s/g}~\left(\frac{t}{1~{\rm day}}\right)^{-1.3},\\
\dot{E}_{\alpha}(t) & \approx&  7\cdot 10^{8}~{\rm
 erg/s/g}~\left(\frac{t}{1~{\rm
	   day}}\right)^{-1}\left(\frac{X_{A\ge 210}}{3\cdot 10^{-2}}\right),\\
\dot{E}_{f}(t) & \approx&  2\cdot 10^{9}~{\rm
 erg/s/g}~\left(\frac{t}{1~{\rm
	   day}}\right)^{-1}\left(\frac{X_{A\ge 250}}{2\cdot 10^{-2}}\right).
\end{eqnarray}
where $X_{A\ge 210}$~($X_{A\ge 250}$) is the total mass fraction of 
nuclei with $210 \le A \le 280$ ($250 \le A \le 280$).  
Note that the numerical coefficients in
Eqs.~(2) and (3) are valid as long as material has the
solar-like $r$-process pattern containing the
second ($A \sim 130$) and third ($A \sim 195$) $r$-process peaks.

Although the form of $\epsilon_{\gamma}(t)$ should be computed with a radiative transfer
simulation, here we give rough estimates.
The optical depth of homologously expanding ejecta 
is given by
\begin{eqnarray}
\tau_{\gamma}(t)= \left(\frac{t_{\rm tr,\gamma}}{t}\right)^{2},\label{tau1}
\end{eqnarray}
where 
$t_{\rm tr,\gamma} \approx (\kappa_{\gamma}M_{\rm ej}/4\pi v^{2})^{1/2}\approx 
0.4~{\rm day}(\kappa_{\gamma}/0.05~{\rm cm^{2}/g})^{1/2}(M_{\rm ej}/0.01M_{\odot})^{1/2}(v/0.3c)^{-1}$ 
is the time that the ejecta become transparent to $\gamma$-rays.
Here we assume that the dominant interaction process of $\gamma$-rays with matter is Compton scattering.

\begin{figure*}
\includegraphics[width=80mm]{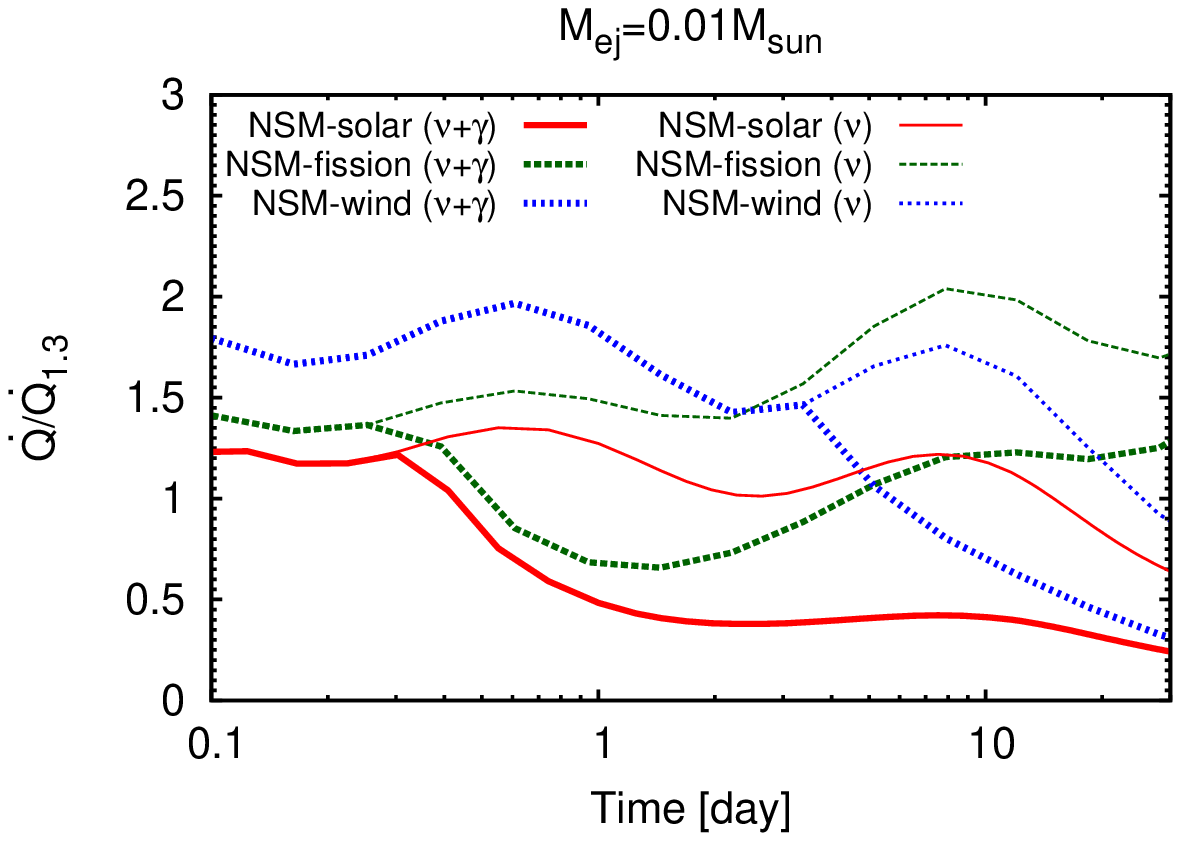}
\includegraphics[width=80mm]{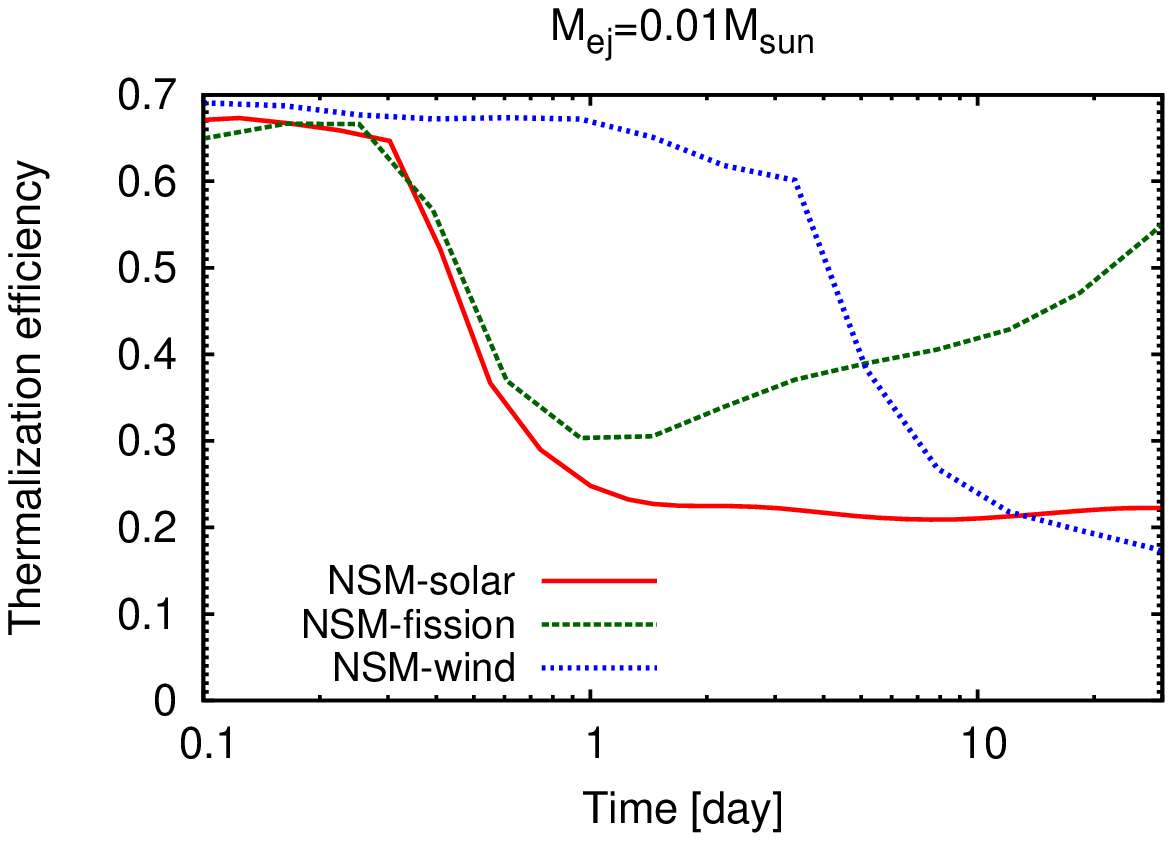}
\caption{Heating rate normalized by $\dot{Q}_{1.3}(t)=10^{10}t_{\rm day}^{-1.3}$~erg/s/g~(left),
where $t_{\rm day}$ is time in unit of day, and thermalization efficiency~(right)
for each model: NSM-solar~(red solid), NSM-fission~(green dashed), and NSM-wind~(blue dotted).
The thick~(thin) lines in the left panel show the heating rate taking 
the neutrino and gamma-ray escape~(only the neutrino escape) into account.
For the gamma-ray escape, an ejecta mass of $0.01M_{\odot}$ is assumed as an example.
}
\label{fig:heat}
\end{figure*}

At the diffuse-out timescale of thermal photons~(optical to infrared: IR) $t_{\rm diff,o}$ 
when the optical depth to thermal photons satisfies $\tau_{\rm opt}=c/v$, 
a significant amount of the deposited energy starts to escape as thermal photons.
We rewrite Eq.~(\ref{tau1}) in terms of  $t_{\rm diff,o}$:
\begin{eqnarray}
\tau_{\gamma}(t) & \approx & \frac{\kappa_{\gamma}}{\kappa_{\rm o}}\frac{c}{v}\left(\frac{t_{\rm diff,o}}{t}\right)^{2},\\
& \approx &  0.02~\left(\frac{t_{\rm diff,o}}{t}\right)^{2}
\left(\frac{\kappa_{\gamma}}{0.05~{\rm cm^{2}/g}}\right) \nonumber \\
& & ~~~~~~~~\times \left(\frac{\kappa_{\rm o}}{10~{\rm cm^{2}/g}}\right)^{-1}
\left(\frac{v}{0.3c}\right)^{-1},
\end{eqnarray}
where $\kappa_{o}$ is the opacity of $r$-process elements to
photons in the optical bands. It is dominated by  bound-bound
transitions of lanthanides~\citep{kasen2013ApJ,tanaka2013ApJ}.
For the dynamical ejecta, on the timescale of $t_{\rm diff,o}$, 
the optical depth to $\gamma$-rays is much smaller than unity, 
thereby only a small fraction of the $\gamma$-rays' energy 
is deposited in the ejecta on the peak timescale of macronovae. 

For the slowly expanding wind ejecta, in particular lanthanide free cases,
the $\gamma$-ray heating efficiency is significantly different.
The opacity to thermal photons and expansion 
velocity of the wind ejecta are 
$\kappa_{o}\sim 1~{\rm cm^{2}/g}$ and $v\sim 0.05c$~(see e.g., \citealt{tanaka2013ApJ} for
the opacity of the wind case). 
The estimated optical depth to $\gamma$-rays is $\tau_{\gamma}\sim 3$ 
on the timescale of $t_{\rm diff,o}$. Therefore, in the case of 
the lanthanide free wind ejecta, $\gamma$-rays are weakly coupled 
with the ejected material and heat up the ejecta until a 
few times $t_{\rm diff,o}$. This situation is somewhat similar to those
of supernovae, in which $\gamma$-rays released from the radioactive decay
of $^{56}$Ni and $^{56}$Co efficiently 
heat up the ejecta on the peak timescale of supernovae~\citep{lucy2005A&A}.

Here we approximately evaluate $\epsilon_{\gamma}(t)$ as the followings.
In optically thick regimes $\tau_{\gamma}\gg 1$, almost all the $\gamma$ rays' energy
is deposited in the ejecta. On the contrary, in optically thin regimes $\tau_{\gamma}<1$, 
only a fraction $\tau$ of the photons are scattered and for each scattered photon
roughly half of $\gamma$ ray's energy is transferred to an electron
via a single scattering process at energies of $\sim 1$~MeV. 
$\epsilon_{\gamma}$ is approximately given by
\begin{eqnarray}
\epsilon_{\gamma}(t) \approx 
\left\{
\begin{array}{ll}
1 - \left(\frac{1}{2}\right)^{N}&~~(\tau_{\gamma}\geq 1),\\
\frac{1}{2}N
&~~(\tau_{\gamma} <1),\\
 \end{array}
 \right.
\end{eqnarray}
where $N=\max(\tau_{\gamma},\tau_{\gamma}^{2})$ is the number of
the scatterings that a photon undergoes before escaping. 
Figure~\ref{fig:heat} shows the heating rate
(left) and the thermalization efficiency~(right). 
Here the heating rate is normalized by a simple power law heating
$\dot{Q}_{1.3}(t)=10^{10}t_{\rm day}^{-1.3}$~erg/s/g~\citep{korobkin2012MNRAS},
where $t_{\rm day}$ is time in unit of day.
The thick~(thin) lines in the left panel show the heating rate taking 
the neutrino and gamma-ray escape~(only the neutrino escape)
into account.
To calculate $\epsilon_{\gamma}(t)$,
an ejecta mass of $0.01M_{\odot}$ is assumed. 
The velocities of the ejecta is set to be $0.3c$ for NSM-solar and NSM-fission
and $0.05c$ for NSM-wind.
For NSM-solar and NSM-fission, at $0.5$~days, 
$\gamma$-rays start to escape from the ejecta so that 
the thermalization efficiency drops from $0.7$ to $0.2$~--~$0.3$.
The heating rate and thermalization efficiency of NSM-wind 
are larger than the other cases since $\gamma$-rays are well scattered. 
For NSM-fission, the thermalization efficiency turns to increase
due to the spontaneous fission of transuranic nuclei
and it reaches $\approx 0.5$ around $10$~days.

\section{Implication to the possible macronova events}
Assuming a constant thermalization efficiency of $0.5$, 
the minimal masses of ejected $r$-process material are
estimated as $\approx 0.02M_{\odot}$ for GRB~130603B~\citep{hotokezaka2013ApJL,piran2014} 
and $\approx 0.1M_{\odot}$ for GRB~060614~\citep{yang2015NatCo}, where
the large lanthanide opacity is taken into account~\citep{kasen2013ApJ,barnes2013ApJ,tanaka2013ApJ}.
These estimates change once we use the thermalization efficiencies
that we obtained here.
The IR detections were done at $7$~days for 
GRB~130603B and at $12$~days for GRB~060614
after the bursts in the GRBs' rest frames. 
For NSM-solar,
the heat is provided only through electrons so that 
the thermalization efficiency at this timescale is $\approx 0.2$~(see Fig.~\ref{fig:heat}).
The minimal required masses are larger than
what previously estimated by a factor of $\approx 2.5$.
As a result, the estimated masses 
are $0.05M_{\odot}$ and $0.25M_{\odot}$ respectively. 
The former may be explained in the context of the black-hole neutron 
star merger ejecta. However, the latter is too large for 
the compact binary merger ejecta.  Therefore, in addition to
$\beta$-decay, other sources of energy injection may be required to explain this
event as a compact binary merger.

The spontaneous fission of transuranic nuclei
can increase the heating rates at later times as discussed earlier. 
For NSM-fission, for which the mass fraction of
transuranic nuclei is $\approx 0.02$,  
the heating rate at $10~{\rm days}$ is larger than that of NSM-solar
by a factor of $\approx 3$~(see the left panel of Fig.~\ref{fig:heat}).
Based on NSM-fission, therefore, 
the minimal required masses are smaller than the ones of NSM-solar by a 
factor of 3 and similar to what estimated in the previous works
\citep{hotokezaka2013ApJL,piran2014, yang2015NatCo}.
However, there are large
uncertainties in the fission and $\beta$-decay lifetimes
of transuranic nuclei~\citep{wanajo2014ApJ} as well as the abundance 
distribution~\citep{eichler2015ApJ}. Thus these estimates are uncertain. 

It is also worthy to mention that there are alternative scenarios to inject additional energy
from the central engine activity, e.g., X-ray irradiation~\citep{kisaka2015}
and magnetar outflows~\citep{fan2013ApJ,metzger2014MNRASb,gao2015ApJ}. 
Because the central engine models predict observable signatures
in the multi-wavelength, to identify the dominant energy source of 
macronovae, X-ray observations on the macronova peak timescale 
and late-time radio observations~(months to years) are important.
In fact, the late time radio upper limits have already constrained
the magnetar engine model for GRB~130603B and 060614~\citep{fong2014ApJ,horesh2015}

\section{Detectability of $\gamma$-rays of $r$-process nuclides}
\begin{figure}
\includegraphics[width=80mm]{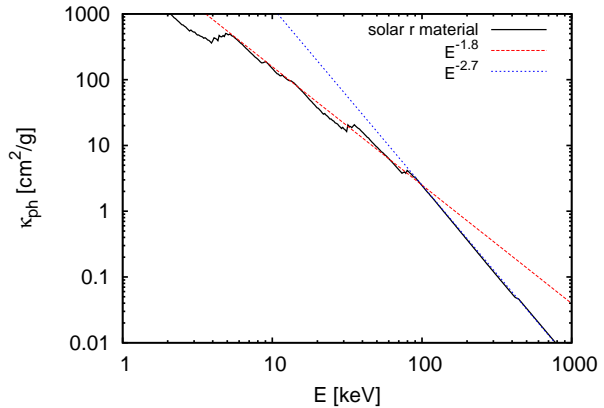}
\caption{Mass absorption coefficient of photoelectric absorption. Here we assume
that the material is composed of $r$-process elements with $40 \leq Z\leq 92$ and the abundance pattern
is the solar pattern of stable and long-lived $r$-process nuclei.}
\label{fig:opacity}
\end{figure}

Given that the $\gamma$-rays escape from the ejecta we turn now to
compute the light curves of $\gamma$-rays 
escaping directly from radioactive decay, i.e.,
$\gamma$-rays produced outside their photosphere~\citep{clayton1969ApJ}.
For these $\gamma$-rays, the observed spectrum preserves the original 
shape~(see Fig.~\ref{fig:spectrum}).
To identify the radius of the photosphere of each energy band,
we take into account three processes
of $\gamma$-rays with matter:
(i) the photoelectric absorption~($E_{\gamma}\lesssim 300~{\rm keV}$),
(ii) Compton scattering~($E_{\gamma }\sim 1$~MeV), and 
(iii) the pair production~($E_{\gamma} \gtrsim 5$~MeV). 
The photoelectric absorption dominates in the low energy range and
depends on the composition of material.
The mass absorption coefficient of the photoelectric absorption for
$r$-process material with the solar abundance pattern~($40\leq Z \leq 92$)
is shown in Fig.~\ref{fig:opacity}.
At high energies $E_{\gamma}\gtrsim 100~{\rm keV}$, 
the mass absorption coefficient $\kappa_{\rm ph}$ is approximately described by
$\kappa_{\rm ph}\approx 2.5~{\rm cm^{2}/g}~(E_{\gamma}/100~{\rm keV})^{-2.7}$. 
At low energies $E_{\gamma}\lesssim 100~{\rm keV}$,  
$\kappa_{\rm ph}\approx 2.5~{\rm cm^{2}/g}~(E_{\gamma}/100~{\rm keV})^{-1.8}$
(see NIST database\footnote{ http://www.nist.gov/pml/data/ffast/index.cfm}
for the mass absorption coefficient for each element).
Here the density structure of ejecta is assumed as $\rho \propto r^{-3}$
and the maximum velocity is set to be twice of the average velocity.

\begin{figure*}
\includegraphics[width=80mm]{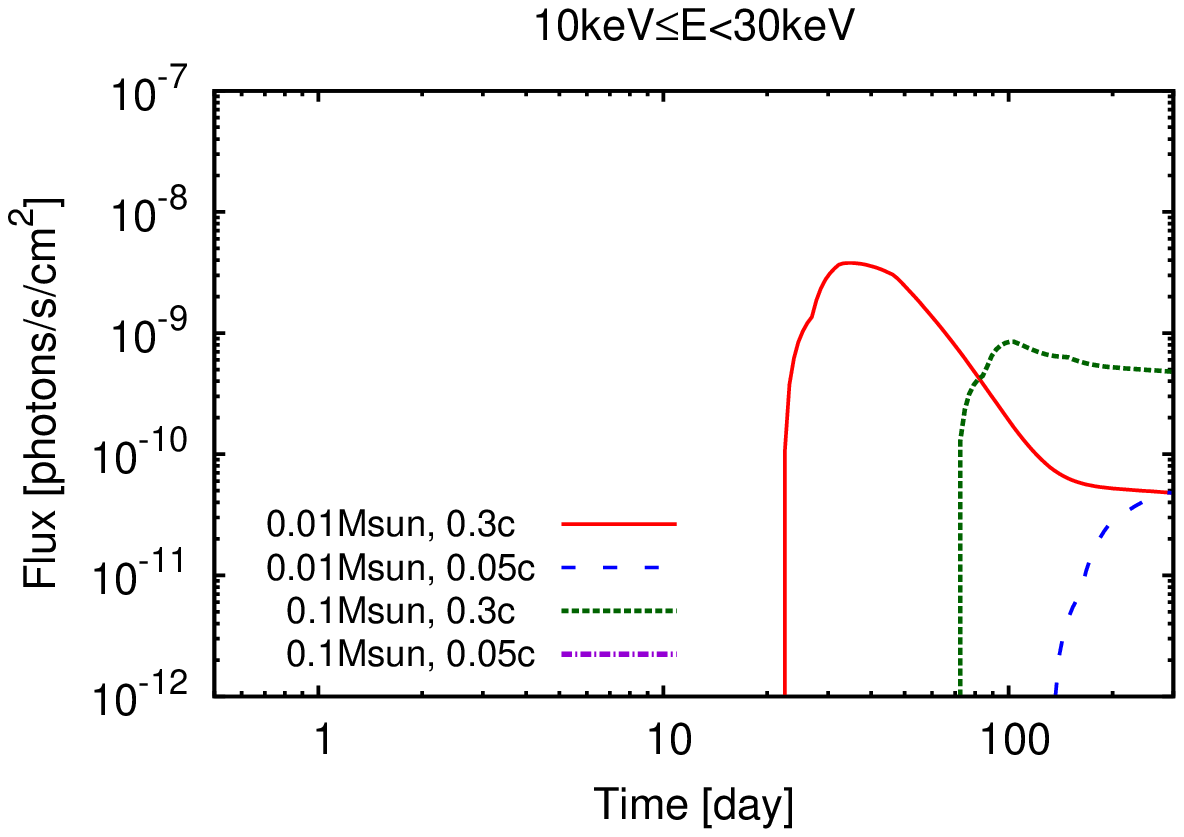}
\includegraphics[width=80mm]{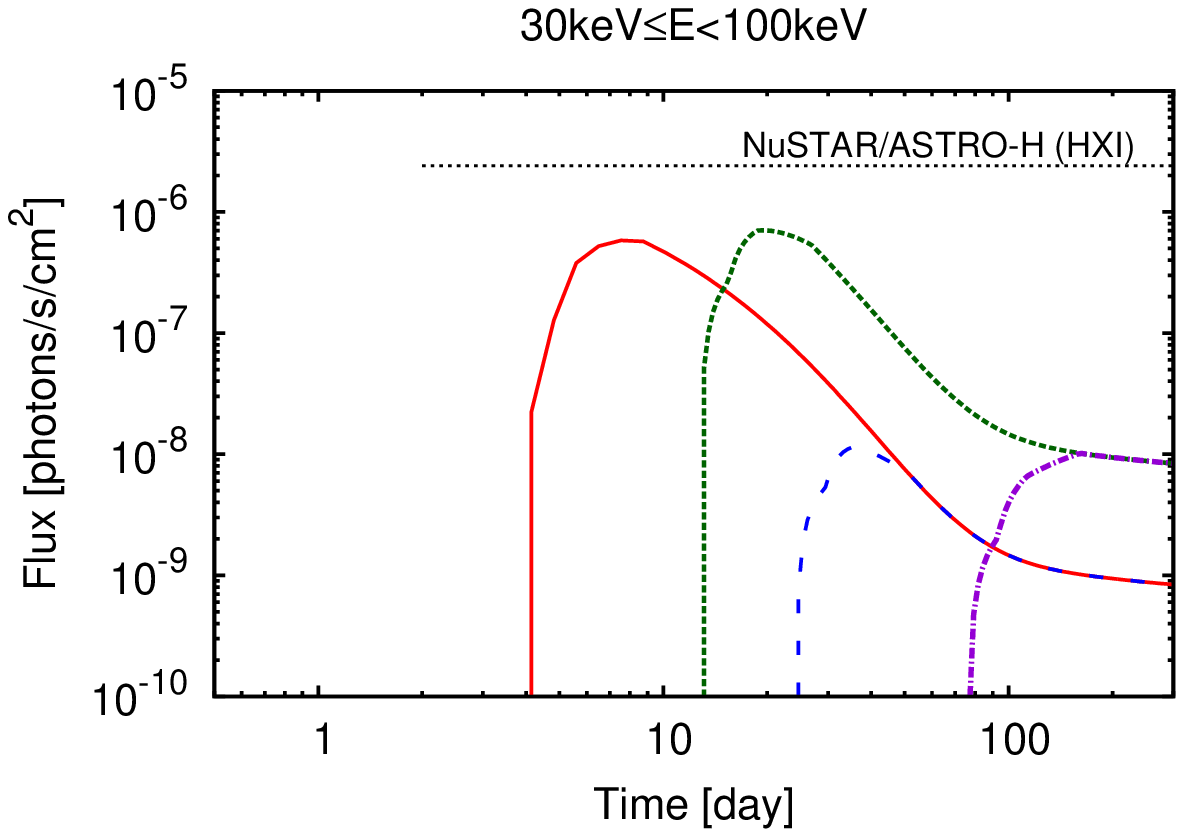}\\
\includegraphics[width=80mm]{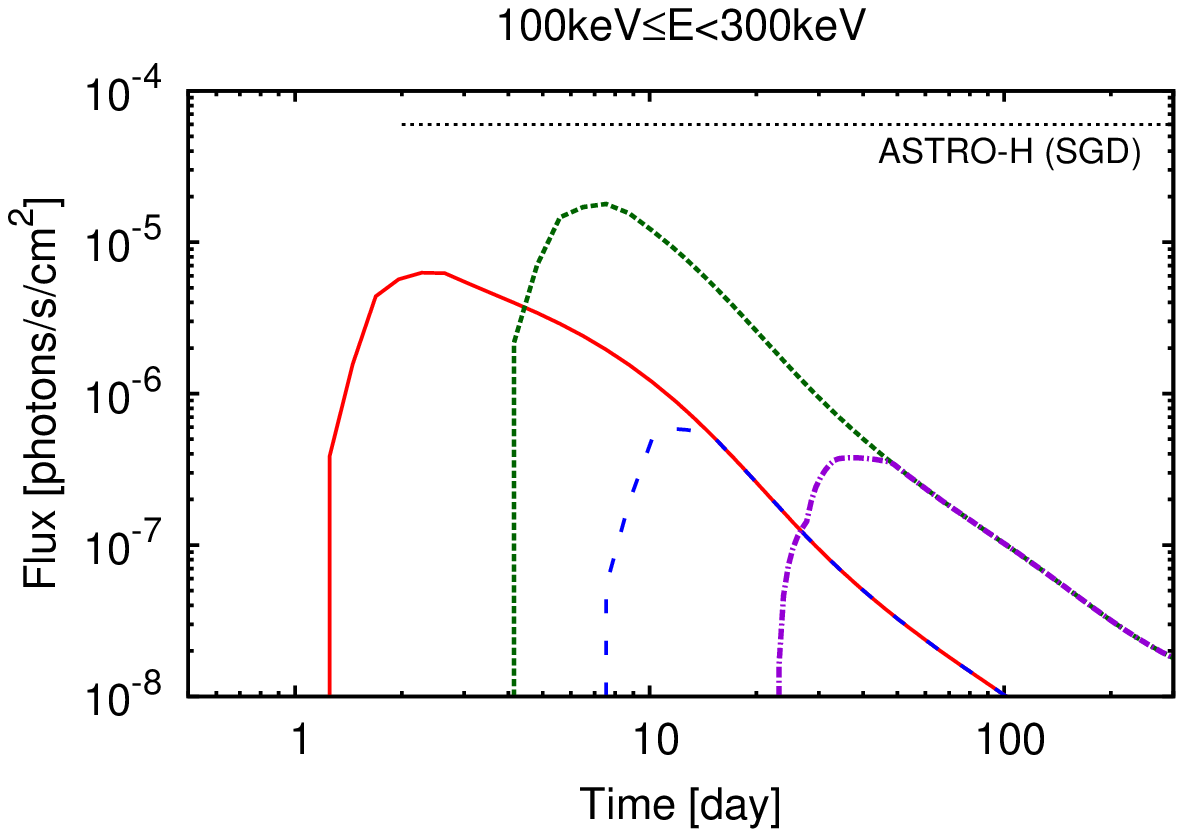}
\includegraphics[width=80mm]{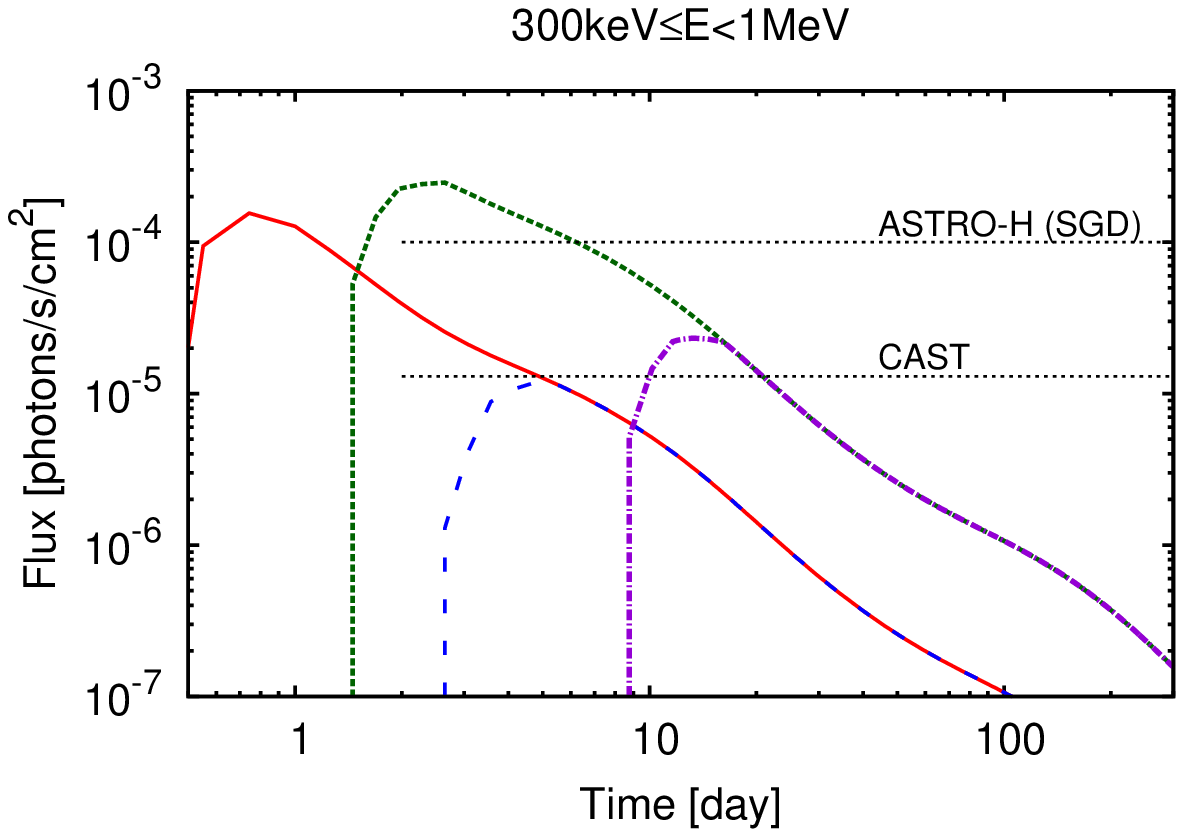}\\
\caption{Light curves of nuclear $\gamma$-rays for NSM-solar 
in the ranges of $10~{\rm keV}\leq E_{\gamma} < 30~{\rm keV}$~(top left),
$30~{\rm keV}\leq E_{\gamma} < 100~{\rm keV}$~(top right),
$100~{\rm keV}\leq E_{\gamma} < 300~{\rm keV}$~(bottom left), and
$300~{\rm keV}\leq E_{\gamma} < 1~{\rm MeV}$~(bottom right) at a distance of $3$~Mpc.
Here we show the four different ejecta models: $(M_{\rm ej},~v)=(0.01~{\rm M_{\odot}},~0.3c)$,
$(0.01M_{\odot},~0.05c)$, $(0.1M_{\odot},~0.3c)$, and $(0.1M_{\odot},~0.05c)$.
Also shown are the sensitivity with exposure at $100$~ks
of current and future X-ray missions: 
{\it NuSTAR}~\citep{nustar2013}, 
{\it ASTRO-H}~\citep{astroh2012},
and {\it CAST}~\citep{cast2014}.}
\label{fig:lightcurve}
\end{figure*}

Figure~\ref{fig:lightcurve} shows the $\gamma$-ray fluxes for an observer at a distance of $3$~Mpc 
in the four different energy bands, $10~{\rm keV}\leq E_{\gamma}<30{\rm keV}$~(top left),
$30~{\rm keV}\leq E_{\gamma} < 100~{\rm keV}$~(top right),
$100~{\rm keV}\leq E_{\gamma} < 300~{\rm keV}$~(bottom left), and
$300~{\rm keV}\leq E_{\gamma} < 1~{\rm MeV}$~(bottom right).
Here we show the four different ejecta models: $(M_{\rm ej},~v)=(0.01~{\rm M_{\odot}},~0.3c)$,
$(0.01M_{\odot},~0.05c)$, $(0.1M_{\odot},~0.3c)$, and $(0.1M_{\odot},~0.05c)$.
Because the photospheric radius depends strongly on the $\gamma$-ray's energy,
the peak timescales of $\gamma$-ray fluxes are different for different energy bands.
For $10~{\rm keV}\leq E_{\gamma} <30~{\rm keV}$, the peak time is as late 
as $30$~days to $100$~days. On the contrary, for $300~{\rm keV}\leq E_{\gamma} <1~{\rm MeV}$, 
the $\gamma$-ray flux peaks around $1$--$10$~days. 
Because more energy is released in $\gamma$-rays at higher energies and 
they escape from the ejecta at earlier times than at lower energies,
the detections may be easier  at the higher energy bands.

Here we compare the expected fluxes of events with the sensitivity of
current-and-future missions in the $30$~keV--$1$~MeV range, i.e.,
{\it ASTRO-H} HXI~(pre-launch) and {\it NuSTAR} for $30$--$100$~keV~\citep{astroh2012,nustar2013},
{\it ASTRO-H} SGD~(pre-launch) for $100$~keV--$1$~MeV~\citep{astroh2012},
and {\it CAST} for $300$~keV--$1$~MeV~\citep{cast2014}.
The resultant sensitivities are shown with the dotted lines Fig.~\ref{fig:lightcurve}
assuming each exposure at $100$~ks, since the expected fluxes show
the variabilities on timescales of $\sim 100$~ks.
$\gamma$-rays are detectable at $300~$keV to $1$~MeV with {\it ASTRO-H} SGD and {\it CAST}
for an event at a distance of $3$~Mpc with an ejecta mass of $0.1~M_{\odot}$.  
For {\it CAST}, it is detectable even for an event at $10$~Mpc with $0.1~M_{\odot}$ and
at $3$~Mpc with $0.01~M_{\odot}$. However, the rate of such nearby events~($\lesssim 3$~Mpc)
is small, e.g., an optimistic estimate gives $\sim 10^{-3}~{\rm yr^{-1}}$~\citep{abadie2010CQG}. 
Therefore, new detectors, more sensitive by at least a factor of ten, are needed for a realistic detection rate
of these signals.

\section{Conclusion}
We studied the radioactive decay products of heavy $r$-process nuclei
in neutron star merger ejecta on timescales from hours to a month. 
We found that $30$--$40\%$ of the energy
is released in neutrinos and lost.
Electrons carry $10$--$20\%$ of the energy. 
These always deposit their energy in the ejecta  and hence this
 provides a lower limit on  the  energy fraction deposited in the ejecta.
In the case that transuranic nuclei
with mass numbers of $A>238$ exist, spontaneous
fission products can carry a significant fraction of the energy  
after a few days. At a week after the merger
this  fraction of the total energy can be $20 \%$ when
the total mass fraction of nuclei with $238<A\leq 280$
is $2\%$. It should be noted, however, the contribution
of spontaneous fission is highly dependent on several
experimentally unknown fission and $\beta$-decay
lifetimes~\citep{wanajo2014ApJ} as well as the
abundance distribution~\citep{eichler2015ApJ}
in this range.

$\gamma$-rays carry $20$--$50\%$ of the energy.  
The number of  $\gamma$-ray photons is roughly
constant per unit energy interval from a few dozen keV
to $1$~MeV. Thus the energy is dominated by $1$~MeV photons. 
The heating efficiency of $\gamma$-rays
depends on the ejecta properties. For the rapidly expanding
ejecta~(e.g. dynamical ejecta), only a small fraction (a few percent) of the 
$\gamma$-rays' energy is deposited at  a few days, that is during  the peak of 
the optical--IR emission. On the contrary,  a large fraction of the energy may be deposited in 
a slowly expanding lanthanide-free ejecta at the corresponding time of the peak.
Such ejecta could be the cases for the late-time wind from 
NSM remnants~\citep{metzger2014MNRAS,perego2014MNRAS,sekiguchi2015PRD,foucart2015} or from
black-hole accretion torii~\citep{wanajo2012ApJ,just2015MNRAS,fernandez2015MNRAS,kasen2015MNRAS} 

Full radiation transfer simulations are needed  in order to obtain the time evolution 
of the $\gamma$-ray energy deposition fraction.
But even without these detailed calculations it is evident that radiation losses of 
the $\gamma$-ray emission will reduce the 
strength of the optical/IR luminosity of the dynamical ejecta, namely the expected macronova 
signature to about two thirds to a half from its original estimates, assuming that this energy 
is absorbed by the ejecta and re-radiated at low energies. For observed signals with a given luminosity 
(e.g. the macronova candidates 130603B~\citealt{tanvir2013Nature,berger2013ApJ} and 060614~\citealt{yang2015NatCo}),
this means that the implied mass should be larger than what was earlier estimated by a factor of two to three.
If a sufficient amount of transuranic nuclei are synthesized, the spontaneous fission products
contribute the heating rate on the 
macronova timescales and the estimated mass can be similar to the previous estimates.
Alternatively, energy injection from the central engine activity can also 
contribute to power the  optical/IR transients~\citep{fan2013ApJ,metzger2014MNRASb,kisaka2015,gao2015ApJ}.

The peak timescale of the flux of the of the escaping $\gamma$-rays
depends significantly on the energy band. For $\gamma$-rays at higher energies of 
$300$~keV to $1$~MeV, the peak timescale is days to a few dozen days.
Due to the photoelectric absorption, the peak time for lower energies of $10$~keV to $30$~keV, is a month to a year. 
The corresponding flux of  $\gamma$-rays at energies of $300$~keV to $1$~MeV
from a merger event with an ejecta mass of $0.1~M_{\odot}$
are detectable with {\it ASTRO-H} SGD and {\it CAST} if it happens at $3$~Mpc and at $10$~Mpc, respectively. 
Direct measurements of these  $\gamma$-rays may provide ultimate proof
of the macronova mechanisms and the sites of $r$-process nucleosynthesis.
However, such a nearby event is too rare to be detected with current detectors. 
This detection with a realistic rate will  have to wait for a more sensitive generation 
that may be launched  in the future.

\section*{Acknowledgments}
We thank Ben Margalit, Satoshi Chiba, and Re'em Sari for discussions.
This research was supported by  the I-CORE Program of the Planning and Budgeting Committee and 
The Israel Science Foundation (grant No 1829/12), the RIKEN
iTHES Project, and Grants-in-Aid for Scientific Research of JSPS
  (26400232, 26400237, 15H02075) and MEXT (15H00788, 15H00773, 15K05107).
\bibliographystyle{mn2e}

\end{document}